\documentclass[final]{svjour3}
\usepackage{graphicx}
\usepackage{rotating}
\usepackage{amssymb}
\usepackage{mathptmx}
\usepackage[numbers]{natbib}
\usepackage{xcolor}
\makeatletter
\journalname{Journal of Low Temperature Physics}

\bibpunct{[}{]}{,}{n}{}{,}

\begin{document}

\newcommand{\hdblarrow}{H\makebox[0.9ex][l]{$\downdownarrows$}-}

\title{Micro-X Sounding Rocket Payload Re-flight Progress}

\author{
Adams,~J.S. \and 
Bandler,~S.R. \and 
Bastidon,~N. \and 
Eckart,~M.E. \and 
Figueroa-Feliciano,~E. \and 
Fuhrman,~J. \and 
Goldfinger,~D.C. \and 
Hubbard,~A.J.F. \and 
Jardin,~D. \and 
Kelley,~R.L. \and 
Kilbourne,~C.A. \and 
Manzagol-Harwood,~R.E*. \and 
McCammon,~D. \and 
Okajima,~T. \and
Porter,~F.S. \and 
Reintsema,~C.D. \and 
Smith,~S.J.
}

\institute{*Corresponding Author\\ Department of Physics and Astronomy, Northwestern University,\\ Evanston, IL 60208, USA\\ 
\email{reneemanzagol@u.northwestern.edu}}

\maketitle

\begin{abstract}
Micro-X is an X-ray sounding rocket payload that had its first flight on July 22, 2018. The goals of the first flight were to operate a transition edge sensor (TES) X-ray microcalorimeter array in space and take a high-resolution spectrum of the Cassiopeia A supernova remnant. The first flight was considered a partial success. The array and its time-division multiplexing readout system were successfully operated in space, but due to a failure in the attitude control system, no time on-target was acquired. A re-flight has been scheduled for summer 2022. Since the first flight, modifications have been made to the detector systems to improve noise and reduce the susceptibility to magnetic fields. The three-stage SQUID circuit, NIST MUX06a, has been replaced by a two-stage SQUID circuit, NIST MUX18b. The initial laboratory results for the new detector system will be presented in this paper. 

\keywords{TES, X-ray, sounding rocket, energy resolution, spectrum, SQUID multiplexing}

\end{abstract}

\section{Introduction}

Micro-X is a rocket-borne X-ray telescope that uses a transition edge sensor (TES) microcalorimeter array to perform high-resolution spectroscopy. The payload is equipped with a Wolter-I imaging optic for focusing X-rays onto the TES array. The array is read out with two Time-Division Multiplexing SQUID readout chains.  Each chain reads out half of the array (``side x" and ``side y") for redundancy in case of a failure to one chain in flight. Temperature control is achieved with an Adiabatic Demagnetization Refrigerator. Refer to Table~\ref{table:DetectorSpecs} for the instrument specifications.

\begin{table}
    \begin{tabular}{ll}
    \hline
		\textbf{Parameter} & \textbf{Specification}\\
		\hline
		Science Observation & ~300 s \\
		Energy bandpass & 0.3 - 2.5~keV \\0
		Microcalorimeter Array & 128 pixels \\
		Field of view & 11.8'\\
		Effective area (1~keV) & 300~cm$^2$\\
		Spectral resolution (0.2 - 3.0~keV) & Goal: 4.5~eV\\
		TES transition temperature & 120~mK\\
	   	Operating temperature & 75~mK $\pm$ 10~ mK\\
		
		Field of view / pixel & 0.98' \\
		Angular resolution (HPD) & 2.4 arcminutes\\
	   	\hline
    \end{tabular}
	\caption{The Micro-X instrument specifications are optimized for high-resolution observations of extended X-ray sources. }
	\label{table:DetectorSpecs}
\end{table}

Micro-X had its first flight on July 22, 2018 from White Sands Missile Range (WSMR) in New Mexico, USA. The goal of the first flight was to take a high-resolution spectrum of the Cassiopeia A supernova remnant. The array and its time-division multiplexing readout system were successfully operated in space, but the payload slowly tumbled for the duration of the flight and no time on-target was acquired due to a failure in the attitude control system. However, X-rays from the on-board calibration source were observed by the detectors. There were significant downtimes across pixels, which is attributed to the SQUID readout being susceptible to the changing direction of the Earth’s magnetic field as the payload spun. A detailed instrument paper (under preparation) will provide details of the first flight. A re-flight has been scheduled for summer 2022. 
For the re-flight, modifications have been made to the payload. This paper covers the major modifications made as well as presents the expected performance for the second flight.

\section{Modifications}
After the first flight, modifications have been made to the detector systems to reduce the susceptibility to magnetic fields and to improve noise. 

Several modifications were focused on reducing the magnetic susceptibility. The three-stage SQUID circuit, NIST MUX06a \cite{Stiehl:2011}, has been replaced by a two-stage SQUID circuit, NIST MUX18b, which has a lower magnetic susceptibility \cite{Reintsema:2019}. In addition, the kinematic mounts that support the TES and interface (IF) chips, which hold the MUX chips, on the detector plane, have been replaced from tungsten carbide balls to sapphire balls, since the tungsten carbide balls were measured to be magnetic. The TES chips, IF chips, and their wiring are surrounded by a superconducting niobium magnetic shield. The shield is composed of a conical body and lid, which are joined by a superconducting lead zipper on top of a spacer. While the spacer should be completely covered, the spacer was swapped from a stainless steel wire to a lead ring, so that in case any of it was exposed, it would still be a fully superconducting shield. 


Micro-X science chain data is saved to on-board flash memory and a subset of data is simultaneously sent to a pair of MV encoders in the rocket telemetry system for real-time transmission to ground, in case the flash memory is not accessible after landing. In order to improve noise, the MV encoder clocks have been synchronized to eliminate observed beat frequencies between the two independent readout systems (``side x" and ``side y") which degraded the first flight performance. Post-flight laboratory testing of the electronics has indicated that synchronizing the clocks will eliminate this noise. Micro-X will be the first program to fly synchronized MV encoders.

\section{Magnetic Field Susceptibility}
The SQUIDs are operated in a flux-locked loop, where the output of the SQUIDs is used by a PI controller to adjust the SQUID feedback, which linearizes the SQUID response. This establishes a locked baseline value from which X-ray pulses can be measured. If the SQUIDs are poorly tuned, there may not exist a stable baseline value; the SQUIDs will become ``unlocked" and unable to read out the pixel. This occurred during the observation period of flight and was caused by the rocket tumbling through Earth's magnetic field. The superconducting magnetic shielding was sufficient to protect the SQUIDs from the magnet in the rocket z-direction, normal to the detector plane. However, the shielding was less effective at shielding the SQUIDs from external magnetic fields in the rocket xy-plane. The flown 3-stage SQUID system has a large second stage SQUID (SQ2) effective area due to the superconducting summing coil which connects the SQ2 to the first stage SQUIDs (SQ1) \cite{Stiehl:2011}. The changing magnetic field at the SQ2 summing coil effectively altered the system operating conditions, creating a drift in the locked baseline value and ultimately resulting in unlocked SQUIDs. For a detailed discussion of this unlocking mechanism see \cite{Fuhrman:2021}.

The MUX18b SQUIDs utilize a two-stage, flux-actuated design that reduces the risk that the system will unlock due to local changes in the magnetic field. The SQUID unlocking mechanism observed in flight 1 requires changing the relationship between the SQ1 and SQ2 stages, so by eliminating the SQ2 stage, this particular unlocking mechanism becomes impossible. While changes in the local magnetic field could affect the SQUID array (SA), the SA has independent magnetic shielding and has not been observed to be susceptible to external magnetic fields in lab testing. Although the MUX18b SQ1 stage has a larger effective area than the MUX06a SQ1 stage, the MUX18b SQ1 stage is still expected to be an order of magnitude lower than the effective area of the MUX06a SQ2 summing coil; therefore the overall magnetic susceptibility of the system has been greatly reduced. Although some locked baseline response to external fields is still expected, the upgraded SQUIDs should not lose lock during flight. 

After first flight, magnetic susceptibility tests were conducted on the flight SQUID system with a Helmholtz coil. The tests were conducted again later when the SQUIDs were replaced. The testing confirmed that external magnetic fields could change the operating conditions of the first flight system and unlock the SQUIDs, and that the SQ2 stage was the most susceptible. Tests were conducted with the TESs superconducting, biased in transition, and normal, confirming that TES response is not required to unlock the SQUIDs. It was also confirmed that the two independent readout chains have inverted baseline responses relative to each other, caused by external fields. This is due to the spatial layout of the MUX chips inside the superconducting shield, as was later verified with a COMSOL model, which will be elaborated on in the upcoming instrument paper. A few additional observations were made that prohibited a more direct comparison to flight data: a strong rate effect when the applied field was changed rapidly and baseline hysteresis after such an effect.

\begin{figure}[htbp]
    \begin{center}
    \includegraphics[width=0.8\linewidth,keepaspectratio]{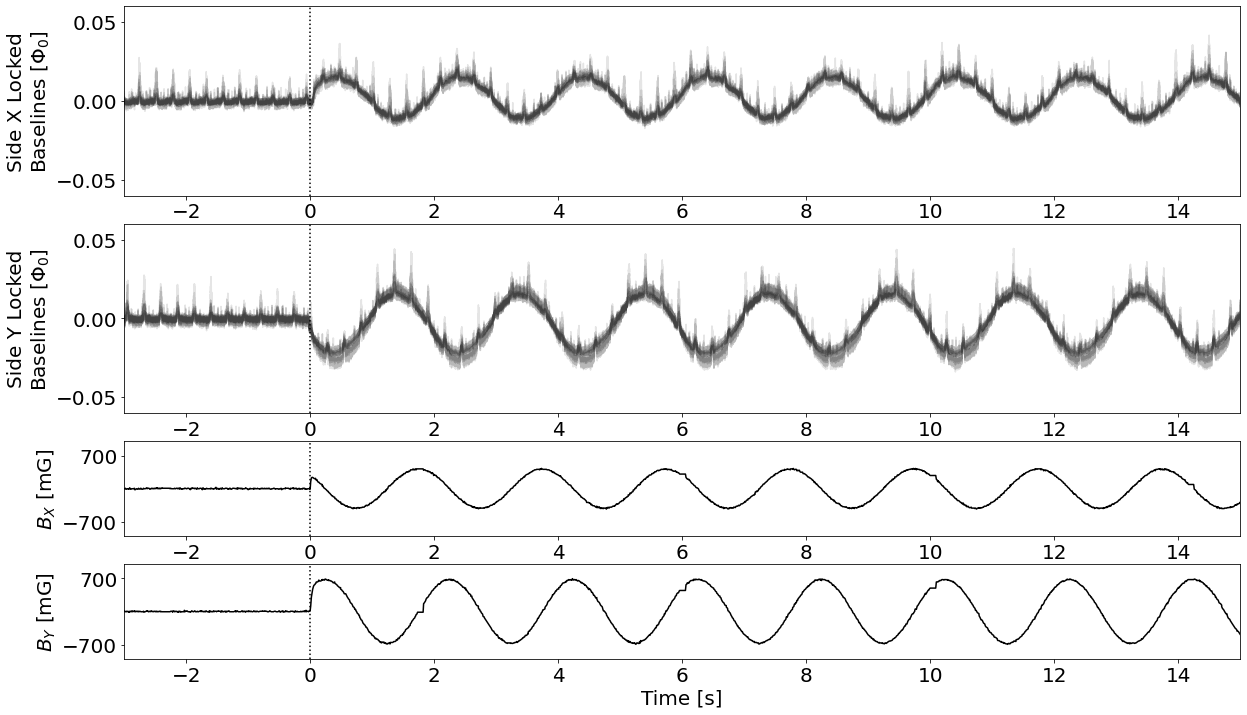}
    \caption{Locked baseline response for 12 pixels on each side driven by a 700 mG magnetic field whose direction is rotating around the rocket x-y plane at 0.5 Hz. The magnetic field was produced by offset sinusoidal currents through a set of two Helmholtz coils in the rocket x- and y-directions. To simulate powered flight the pixels were tuned at their operating temperature, 75 mK, and tested at their launch temperature, 300 mK. The repeated coincident baseline spikes were caused by the asynchronous clocks when running both detector sides simultaneously. The DC baselines were set to 0 to overlay the pixel responses. The 3-axis magnetometer was placed on the outside of the cryostat such that $B_x$ could only be measured at the edge of the field produced by the Helmholtz coils, which subsequently under-measured the field in the x-direction. Time is referenced from when the coils are powered on.}
    \label{fig:helmholtz}
    \end{center}
\end{figure}

Helmholtz coil tests confirm the significantly reduced magnetic susceptibility. The SQUIDs remained locked under $\pm 1.8 B_{Earth}$ when applied along the 3 primary axes. No rate or significant hysteric effects have been observed. The baseline response from a test using two Helmholtz coils to apply an external field that rotates about the thrust axis, to simulate powered flight, is shown in Fig. \ref{fig:helmholtz}. During powered flight, the rocket spins at an average of 4 Hz about its thrust (z) axis for stability, lasting approximately 62 seconds. The changing orientation of the magnetic field can create a baseline response like what was observed in the first flight. The baseline oscillations measured 3.3e-5~$\Phi_0$~per~mG applied, where $\Phi_0$ is the magnetic flux quantum. In flight, the MUX06a SQUIDs exhibited the same behaviour at 1.1e-4~$\Phi_0$~per~mG. If the magnetic shield performance is comparable between both cases, the MUX18b SQ1 effective area is approximately 80 $\mu m^2$, significantly reduced from the 468.5~$\mu m^2$~SQ2~effective~area of the MUX06a SQ2 \cite{Stiehl:2011}. While susceptibility to external fields is always a risk with TESs and SQUIDs, it is no longer considered a likely failure for future flights.

\section{Detector Performance}

The primary benchmark of the payload is energy resolution. For detector testing, a calibration source is used to collect X-rays in the science energy region of interest. This section will compare the current performance of the detectors after post-flight modifications to the performance from laboratory testing immediately before the first flight.

Pre-flight data was taken on the ground at WSMR two weeks prior to first flight, but after the instrument had been fully built up in flight configuration. This data was taken in high-speed mode, which has the bandwidth to operate with both science chains on, including independent power supplies and master clocks. The calibration source used was an internal Fe-55 ring that decays to Mn and fluoresces a KCl ring. This gives strong Cl-K$\alpha$ and K-K$\alpha$ peaks at 2622~eV and 3313~eV, respectively. X-ray pulses were triggered with 50~mV edge trigger and noise was triggered with auto trigger every 4096~ms, resulting in roughly half as many noise traces as pulse traces per pixel.

The post-flight data shown here was taken in a laboratory at Northwestern University (NU) almost two years after first flight, in as close to flight configuration as possible. One large difference from the pre-flight data is that this was taken with just one of the science chains operating at a time. One-sided operations at NU replicate the MV encoder clocks being synchronized for reflight.
The calibration source used for this data was an external Fe-55 source directly illuminating the detector array through a Be window, providing a strong Mn-K$\alpha$ peak at 5894~eV. X-ray pulses were triggered with an edge trigger of 0.15~mV/s and noise was triggered with a level threshold of 1000~mV, resulting in roughly five times as many noise traces as pulse traces per pixel.
 Record lengths are 4096 digital samples with 1024 digital samples saved pre-trigger.

\begin{figure}[htbp]
    \begin{center}
    \includegraphics[width=0.8\linewidth,keepaspectratio]{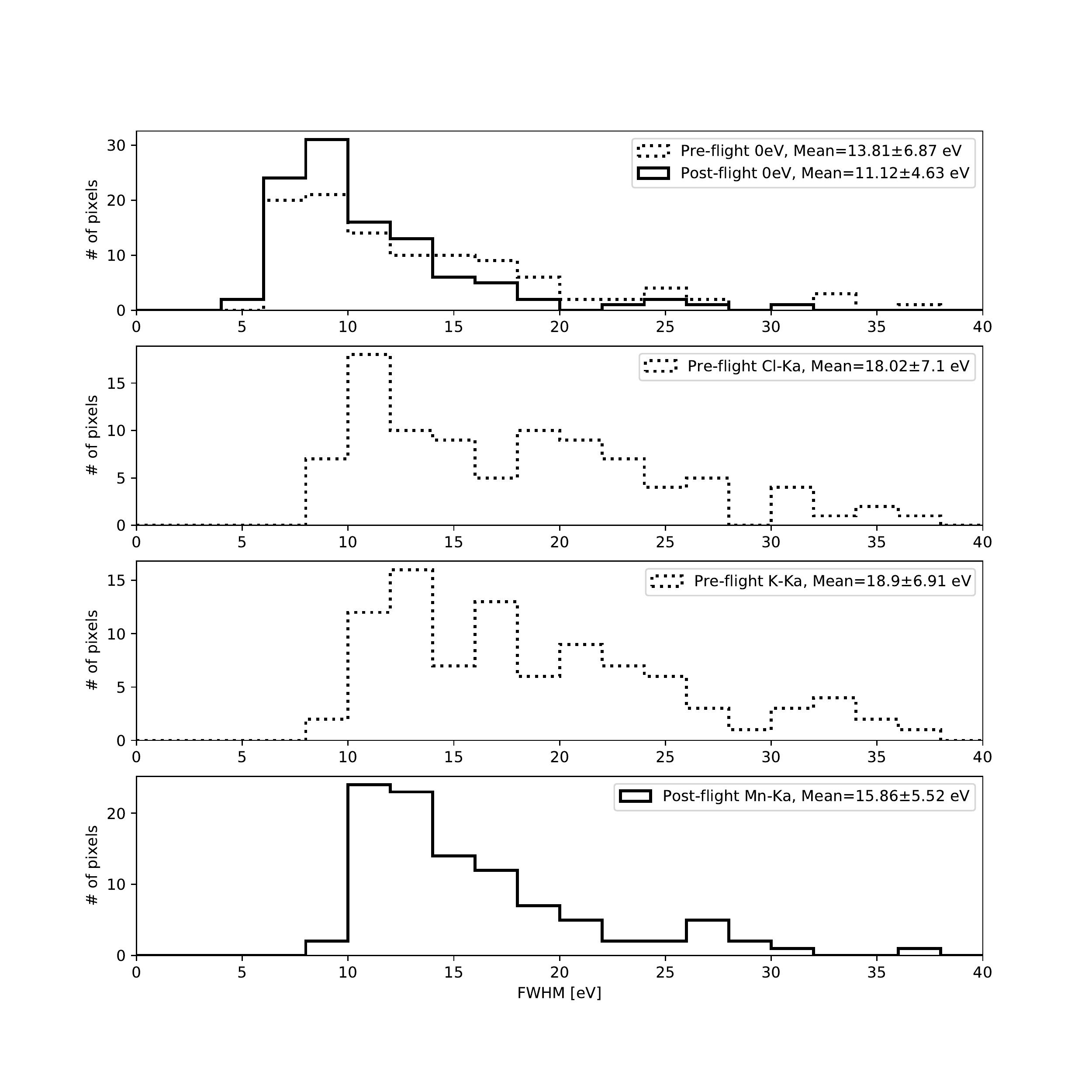}
    \caption{Histograms of the spectral resolution of each peak of the calibration source. The {\it dotted} lines are pre-flight data (taken in 2018) and the {\it solid} lines are post-flight data taken with the new SQUID system in 2021.}
    \label{fig:resolution_hists}
    \end{center}
\end{figure}

The data was analyzed using an Optimal Filter algorithm \cite{XCalGSE}. The triggered data was loaded, and timing and threshold cuts were used to select pulses and noise as well as reject doubles and pileup events. A series of corrections was also applied to the data to correct for  baseline or temperature drifts. Finally, each peak was fit in order to calibrate the gain as a function of energy. All of these corrections were performed manually on each pixel individually. The result is a calibrated energy spectrum where the full-width half-maximum (FWHM) of each peak can be extracted from a Gaussian fit for the 0~eV baseline noise or a Lorentzian fit for the spectral lines.
Fig.~\ref{fig:resolution_hists} shows histograms of the measured resolutions of each peak in both datasets.

\begin{figure}[htbp]
    \begin{center}
    \includegraphics[width=0.6\linewidth,keepaspectratio]{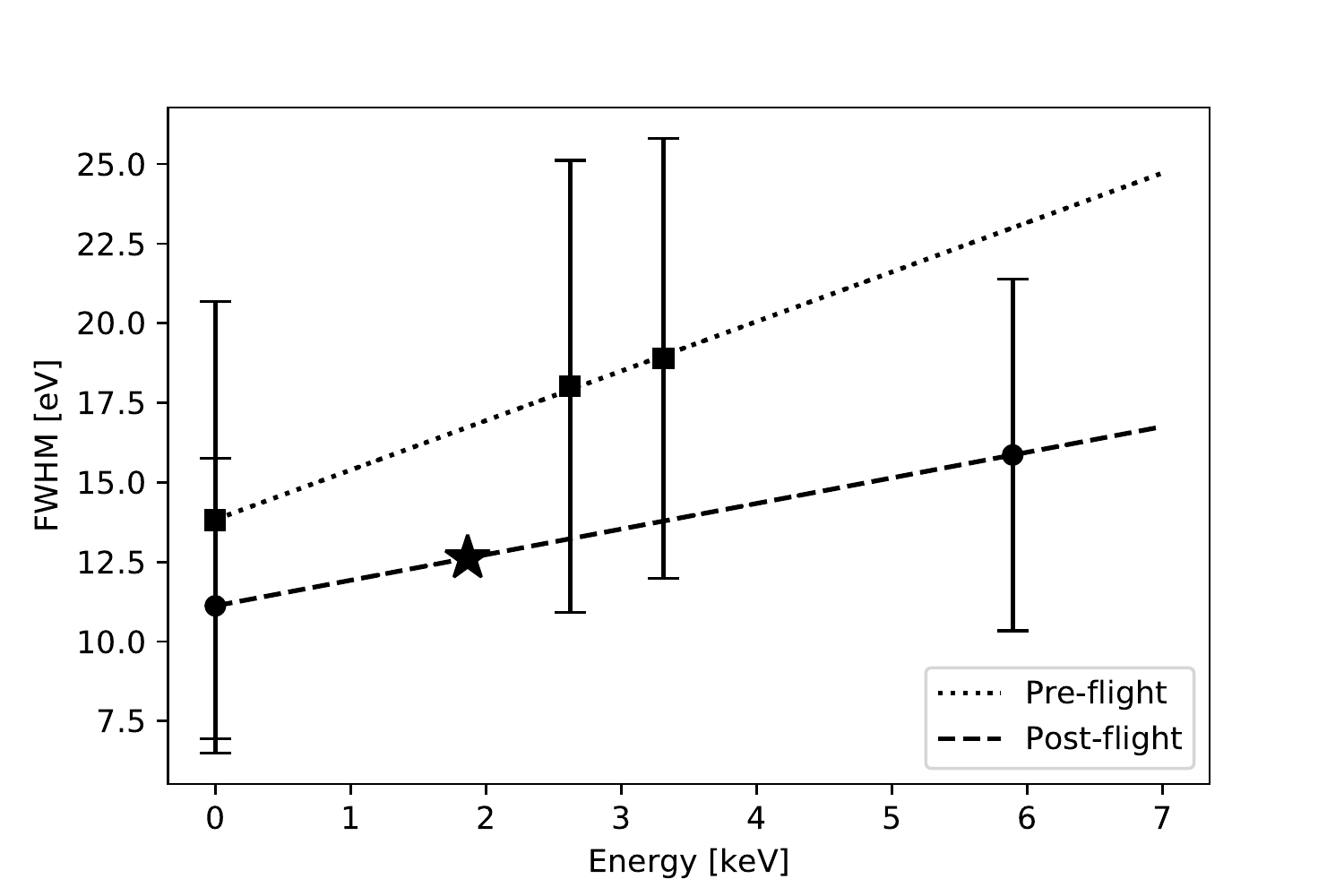}
    \caption{Plot of the mean resolutions of each peak in the pre-flight and post-flight datasets. The {\it errorbars} show the error on the mean, the {\it dotted} line is a linear fit to the pre-flight resolutions, the {\it dashed} line is a linear fit to the post-flight resolutions, and the {\it star} indicates the expected energy resolution at the strongest line of the science target for the next flight.}
    \label{fig:resolution_vs_energy}
    \end{center}
\end{figure}

Fig.~\ref{fig:resolution_vs_energy} shows the mean of each peak resolution as a function of energy for both datasets. The strongest line from the Cas A target is Si~XIII, which has an energy of 1.865~keV. By interpolating a linear fit of the resolutions, the expected energy resolution at 1.865~keV is estimated to be 12.62~eV, indicated by the {\it star} on the plot. This is an improvement over 16.73~eV at 1.865~keV prior to first flight, although this is just within the error bars and can mostly be attributed to the difference between two-sided vs. one-sided modes.

\section{Conclusion}
The first flight of Micro-X marked the first operation of TES arrays and multiplexed SQUID readouts in space. Although the attitude control system failure led to no time on target, it allowed for identification of improvements to the payload for future flights, specifically around magnetic field susceptibility and noise reduction. Laboratory tests of the improved system show reduced magnetic susceptibility and, in the closest flight configuration that can be replicated in the laboratory, show that the new system is surpassing first flight integration performance.

The datasets generated during and/or analysed during the current study are available from the corresponding author on reasonable request.

\begin{acknowledgements}
The Micro-X project is conducted under NASA grant 80NSSC20K0430. Part of this work was performed under the auspices of the U.S. Department of
Energy by Lawrence Livermore National Laboratory under Contract DE-AC52-07NA27344.
R. Manzagol-Harwood acknowledges financial support from the IDEAS Fellowship, a research traineeship program funded by the National Science Foundation under
grant DGE-1450006 and from the Illinois Space Grant Consortium Graduate Fellowship Program.
\end{acknowledgements}

\pagebreak


\begin{thebibliography}{99}

\bibitem{Stiehl:2011}
G.M. Stiehl et al., \textit{IEEE Transactions on Applied Superconductivity}, \textbf{21} 3, (2011), 10.1109/TASC.2010.2091483.

\bibitem{Reintsema:2019}
C. D. Reintsema et al., \textit{IEEE Transactions on Applied Superconductivity}, \textbf{29} 5, (2019), 10.1109/TASC.2019.2904594.

\bibitem{Fuhrman:2021}
J. S. Adams et al., \textit{J. Low Temp. Phys.}, This Special Issue (2021)

\bibitem{XCalGSE}
XCal GSE v11.1.6 [software], NASA/GSFC (2019)






\end{thebibliography}
\end{document}